\pdfoutput=1
\documentclass[aps,pra,10pt,twocolumn,showpacs,superscriptaddress,floatfix]{revtex4-1}

\usepackage[usenames,dvipsnames,svgnames]{pstricks}
\usepackage{newlfont}
\usepackage{amssymb}
\usepackage{amsfonts}
\usepackage{amsmath}
\usepackage{amsthm}
\usepackage{graphicx}
\usepackage{color}
\usepackage[breaklinks]{hyperref}

\usepackage{times}

\begin{document}
\title{Probing Uncertainty Relations in Non-Commutative Space}
\author{Pritam Chattopadhyay}
\affiliation{Cryptology and Security Research Unit, R.C. Bose Center for Cryptology and Security,\\
Indian Statistical Institute, Kolkata 700108, India}
\email{pritam.cphys@gmail.com}

\author{Ayan Mitra}
\affiliation{Energetic Cosmos Laboratory, Nazarbayev University, Nur-Sultan 010000, Kazakhstan}
\email{ayan.mitra@nu.edu.kz\\
Part of this work was done when the second author was visiting R. C. Bose\\ Centre for Cryptology and Security of Indian Statistical Insitute as a Post-Doctoral Researcher during 2017\\}

\author{Goutam Paul}
\email{goutam.paul@isical.ac.in}
\affiliation{Cryptology and Security Research Unit, R.C. Bose Center for Cryptology and Security,\\
Indian Statistical Institute, Kolkata 700108, India}

\pacs{}

\begin{abstract}
In this paper, we compute uncertainty relations for non-commutative space and obtain a better lower bound than the standard one obtained from  Heisenberg's uncertainty relation. We also derive the reverse uncertainty relation for product and sum of uncertainties of two incompatible variables for one linear and another non-linear model of the harmonic oscillator. The non-linear model in non-commutating space yields two different expressions for Schr\"odinger and Heisenberg uncertainty relation. This distinction does not arise in commutative space, and even in the linear model of non-commutative space.
\end{abstract}

\maketitle

\section{Introduction}
\label{intro}

In the early days of quantum field theory, Heisenberg suggested that one can make use of non-commutative structure for spacetime coordinates for small length scales which introduces an effective ultraviolet cutoff. It was Synder~\cite{synd} who first proposed a formalism on this area to control the
divergences which had troubled theories like quantum electrodynamics. Similar to the case of quantization of classical phase space, we can define the non-commutative space by substituting the spacetime coordinates  $x^i$ by their hermitian operators $\hat{x^i}$ of non-commutative $C^{*}$-algebra of spacetime~\cite{gl}. A tactile example of physics in non-commutative spacetime is Yang-Mills theory on non-commutative torus~\cite{mar}. Connes-Lott model explores the physical application of standard model and its various fields which was based on geometric interpretation of this space structure~\cite{jcv,cpm}.  Other quantum field theory~\cite{fli} and gravity~\cite{hch} were introduced along this lines.

The quantum mechanical uncertainty principle (UP) (also called the Heisenberg's uncertainty principle~\cite{heisenber}) in simple term states that no two incompatible variables can be determined simultaneously with an accuracy greater than some fundamental constant. Mathematically, this is given in terms of the standard deviations of the pair of variables~\cite{Weyl}. For two canonical observables $x$ and $p$, it is defined as
\begin{eqnarray}\label{sun0}
\hspace{-.2cm}\Delta x\Delta p\geq\frac{\hbar}{2},
\label{e1}
\end{eqnarray}
However, we would like to clarify two closely associated terms, namely: uncertainty principle(UP) and the uncertainty relation (UR). The UP as stated in~\cite{robert,Schrodinger,Schrodinger1,maccone} and also as stated above, points to the impossibility of the joint measurement of any two incompatible observable pairs. On the other hand the UR doesn't refer to measurement-induced disturbances, rather it refers to the state induced spread in the measurement outcome. In this paper, we will be dealing with the UR.  Starting from Heisenberg's UR~\cite{heisenber}, there have been further new forms developed over the years.  The most common form of the UR was given by Robertson~\cite{robert} as
\begin{eqnarray}\label{sun}
\hspace{-.2cm}\Delta A\Delta B\geq\left|\frac{1}{2}\langle[A,B]\rangle\right|,
\end{eqnarray}
where $A$ and $B$ are two observables. Eq.~\eqref{e1} can be reproduced from the Robertson relation by substituting the corresponding commutation relation for $[x, p]$. Schr\"{o}dinger added an extra anti-commutator term~\cite{Schrodinger,Schrodinger1}, thus further strengthening the  bound 
\begin{eqnarray}\label{b1}
\hspace{-.2cm}\Delta A^2\Delta B^2\geq\left|\frac{1}{2}\langle[A,B]\rangle\right|^2+\left|\frac{1}{2}\left\langle\{A,B\}\right\rangle-{\langle}A{\rangle}{\langle}B{\rangle}\right|^2.
\end{eqnarray}
Further recently, Pati and Maccone~\cite{maccone} showed that there exists a stronger UR, called Pati-Maccone UR (PMUR), with a tighter lower bound. Using the algebraic square of sums ($(a{\pm}b)^2=a^2+b^2{\pm}2ab$), they turned the product form of the uncertainty relation into an additive form:
\begin{eqnarray}\label{sunnn}
\hspace{-.2cm}\Delta A^2+\Delta B^2\geq{max\left(\mathcal{L}_1,\mathcal{L}_2\right)},
\end{eqnarray}
where $\mathcal{L}_1$ and $\mathcal{L}_2$ are defined as 
${\pm}i\langle[A,B]\rangle+\left|\langle{\psi}|A{\pm}iB|{\psi}^\perp\rangle\right|^2$ and
$\tfrac12|\langle\psi_{A+B}^\perp|A+B|{\psi}\rangle|^2$ respectively.
Here $|\psi_{A+B}^\perp\rangle$ is the state, orthogonal to the state of the system  $|\psi_{A+B}\rangle$. We should choose the sign such that ${\pm}i\langle[A,B]\rangle$ yields a positive number.

The PMUR provides a non-trivial solution to the lower bound, which in the previous UR's was missing. For example, in relation ~\eqref{b1}, if $A$ and $B$ are incompatible on the states of the system $|\psi\rangle$, then the whole relation reduces to a trivial case. However, for PMUR, the lower bound is almost always non-trivial (i.e., non-zero) for both the cases where  $|\psi\rangle$ is  a common eigenstate of  $A$ and $B$, and when it is not. The work~\cite{jonas} has made a thorough derivation of PMUR relations, while the work~\cite{kunwang} has given an experimental validation of the PMUR relation. It is easy to see that the previous UR's can be shown as special cases of  the PMUR relation.  

The importance of UR is undeniable in almost all branches of physics and the recent works~\cite{barato,zhang,hyeon,jia,feng,scard,bojo,singh,guo,sch} convey it's importance, especially those involving experiments in quantum domain. Recently, various experimental tests have been performed to verify the UR's~\cite{qmref,fei1,baek}.

In this paper, we also present the analysis using reverse UR, which has been formulated in~\cite{pati}.  This is also useful in capturing the essence of the quantum uncertainties~\cite{deutsch,huang,jorge,jorge1,puchala}. They are the main tool necessary for formulating quantum mechanics~\cite{qmref2,qmref,busch}  and even quantum gravity~\cite{qgref1}.  Technologically, in present time, it is even more important, as it has applications in quantum cryptography~\cite{fuchs,qcref1,qcref2}, and also in quantum entanglement detection ~\cite{qe1,qe2,qe3,guhne}.  It is also used in quantum metrology~\cite{seth} and quantum speed limit research~\cite{qmm1,qmm2,qmm3,mandel,deba,gerardospeed}. Likewise it is also used in space-time~\cite{stref1} and gravity analysis~\cite{stref2}. It has important relevance in string theory~\cite{kempfstring} as well.

Here we will present our results on deriving the UR's from Schr{\"o}dinger's expression Eq.~\eqref{b1} and also the PMUR, both in product and sum of variance forms  in non-commutative space. 
Our choice to solve the complete analysis of~\cite{pati} in a similar way but in a non-commutative space is motivated by the fact that the UR's scenario in NC space were strongly fueled by it's mathematical background. Existing frameworks of modern classical geometry is outlined by Riemann's hypotheses of geometry~\cite{ano}, defined by two important concepts of: manifold and line elements~\cite{R2,R}. The validity of the infinitely small line element is connected to the basis of the respective metric of the space~\cite{ano}. However in quantum mechanics, the domain of the space being operated on fails to be a manifold. 

Heisenberg noted that in a quantum mechanical system (e.g., transition of energy levels in an atom), the product of such observables $A$ and $B$ are not commutative, i.e.,
$AB{\neq}BA$,(e.g, $A$ or $B$ could be the emission or absorption lines of atomic spectra). For such systems, it is better to think of a new kind of space in which the coordinates do not commute. In addition to this, there are abundance of	example spaces which are governed by non-commutative algebra and with obvious relevance to physics and mathematics. The phase space of quantum mechanics is one of the most important one.

Another very important domain for probing with non-commutative algebra is space-time models~\cite{connes2,kalau,connes3,connes4} originating from non-commutative algebra.

This paper is structured into two parts. In the first section, we are going to develop the UR from the Robertson- Schr\"odinger relation for the NC space. We will optimize the uncertainty relation using the same approach used by~\cite{pati} and show that it is stronger than the Robertson-Schr{\"o}dinger UR. In the second section, we will develop the reverse UR using~\cite{pati}, and develop the tighter uncertainty for both the sum and the product of two incompatible observables in NC space.

An interesting result comes out for the non-linear model in NC space. It yields two different expression for Schr\"odinger and Heisenberg UR. The identification for this two relation does not arise in commutative space, and even in linear model of NC space.

\section{Tighter Uncertainty relation in Non-Commutative Space for Linear Harmonic Oscillator (LHO)}
 Till date, various NC space models have been proposed and analyzed. The proposed linear models can be generalized as:
 \begin{equation}\label{sun11}
 X_i= A_{i,j}	x_j,
 \end{equation}
 where $X_i= [X_{1}, X_{2}, \ldots, X_{2n}]^T$,
$x_i= [x_{1}, x_{2}, \ldots, x_{2n}]^T$, and

$A_{i,j}=
\begin{bmatrix}
    a_{1,1} & a_{1,2} & a_{1,3} &\hdots \hdots  a_{1,2n} \\
    a_{2,1} & a_{2,2} & a_{2,3} &\hdots \hdots  a_{2,2n} \\
    a_{3,1} & a_{3,2} & a_{3,3} &\hdots \hdots   a_{3,2n} \\
    \vdots &\vdots &\vdots & \vdots   \\
    a_{2n,1} & a_{2n,2} & a_{2n,3} &  \hdots \hdots a_{2n,2n}.
\end{bmatrix}$

Here $X_j$ are the co-ordinates of the non-commutative phase space and $x_j$ are the known commutative space co-ordinates. The variables $a_{ij}$ can take any constant values. Any linear model can be easily procured from the generalized form given in Eq.~\eqref
{sun11} by replacing the values of the components of $A_{ij}$. In this paper we will present two NC models, (one linear and the other, non-linear). The models are presented below.

\subsection{Model 1: Linear model}
We take
\begin{align}\label{sun15}
\hat{X_1}=\hat{x_1}-\frac{\lambda}{2}\hat{p_2}, && \hat{X_2}=\hat{x_2}+\frac{\lambda}{2}\hat{p_1},\\ \nonumber \hat{P_1}=\hat{p_1}+\frac{\gamma}{2}\hat{x_2}, && \hat{P_2}=\hat{p_2}-\frac{\gamma}{2}\hat{x_1},
\end{align}
where $\lambda$ and $\gamma$ are constants (extracted from the matrix $A_{ij}$ for the current model, where $n=2$.) This model are being used for the analysis of quantum gravity.
The commutation relation for this observables can be written as
\begin{align}
[\hat{X_1},\hat{X_2}]=i\lambda\hbar, \indent
[\hat{P_1},\hat{P_2}]=i\gamma\hbar, \\
[\hat{X_1},\hat{P_1}]=[\hat{X_2},\hat{P_2}]=i\hbar\Big(1+\frac{\lambda\gamma}{4}\Big). \nonumber
\end{align}

\textit{The Robertson-Schr\"odinger uncertainty relation} (RS) for position ($X$) and momentum ($P$) operators are represented as
\begin{equation}\label{sum1}
\Delta \hat{X}^2 \Delta \hat{P}^2\geq \Big|\frac{1}{2}\langle[\hat{X},\hat{P}]\rangle\Big|^2+\Big|\frac{1}{2}\langle\{\hat{X},\hat{P}\}\rangle-\langle\hat{X}\rangle\langle\hat{P}\rangle\Big|^2.
\end{equation}
The expectation values, commutation and anti-commutation relations for the position $X_1$ and the momentum $P_1$ operators for LHO with respect to our known canonical variables are
\begin{equation}\label{commu1}
[\hat{X_1},\hat{P_1}]=i\hbar\Big(1+\frac{\lambda\gamma}{4}\Big),
\end{equation} 
\begin{equation}\label{commu2}
\langle\hat{X_1}\rangle\langle\hat{P_1}\rangle
=\langle\hat{x_1}\rangle\langle\hat{p_1}\rangle+\frac{\gamma}{2}\langle\hat{x_1}\rangle\langle\hat{x_2}\rangle-\frac{\lambda}{2}\langle\hat{p_2}\rangle\langle\hat{p_1}\rangle-\frac{\lambda\gamma}{4}\langle\hat{p_2}\rangle\langle\hat{x_2}\rangle,
\end{equation}
and
\begin{equation}\label{commu3}
\{\hat{X_1},\hat{P_1}\}=\{\hat{x_1},\hat{p_1}\}-\frac{\lambda}{2}\{\hat{p_2},\hat{p_1}\}+\frac{\gamma}{2}\{\hat{x_1},\hat{x_2}\}-\frac{\gamma\lambda}{4}\{\hat{p_2},\hat{x_2}\},
\end{equation}
respectively.
 Plugging in  Eq.~\eqref{commu1},~\eqref{commu2},~\eqref{commu3} in Eq.~\eqref{sum1} we get  
\begin{eqnarray}
\Delta \hat{X_1}^2 \Delta \hat{P_1}^2\geq \frac{\hbar^2}{4} \Big(1+\frac{\lambda\gamma}{4}\Big)^2.
\end{eqnarray} 
So we can see that the RS inequality yields the same results as we can develop from the Heisenberg UR relation ~\eqref{sun0}. For the commutative space, there is no uncertainty in the position or momentum operator. But in NC space, the uncertainties in the position and momentum operators appear naturally.

The uncertainty of position operator can be generated from Eq.~\eqref{b1} just by replacing it with $X_1$ and $X_2$.
Plugging in the commutation, anti-commutation and the expectation relation in Eq.~\eqref{b1} for its corresponding $X_1$ and $X_2$ form, we get 
\begin{equation}
\Delta \hat{X_1}^2 \Delta \hat{X_2}^2\geq \frac{\hbar^2 \lambda^2}{4}.
\end{equation}

Similarly, for the momentum operator, we can develop the uncertainty relation for the NC space by replacing the variables $A$ and $B$ by $P_1$ and $P_2$ in Eq.~\eqref{b1}. 
Using the same formalism as above, we get 
\begin{equation}
\Delta \hat{P_1}^2 \Delta \hat{P_2}^2\geq \frac{\hbar^2 \gamma^2}{4}.
\end{equation}

So we can conclude that the RS relation is equivalent in nature to the well known Heisenberg relation for LHO.

Now we proceed to compute the \textbf{tighter UR} relation. The expression is given as 
\begin{equation}\label{a2}
\Delta A^2 \Delta B^2\geq \max_{\{|\psi_{n}\rangle\}}\frac{1}{4} \Big(\sum_n \Big|\langle[\bar{A},\bar{B_n^\psi}]\rangle_\psi+\langle\{\bar{A},\bar{B_n^\psi}\}\rangle_\psi\Big|^2\Big).
\end{equation}
Eq.~\eqref{a2} is tighter  from Eq.~\eqref{b1} in the sense that it is achieved by optimizing the UR over the complete orthonormal bases. 
We want to optimize our UR in NC space using Eq.~\eqref{a2}.

For the ($X_1$,$P_1$) pair, it reduces to
\begin{equation}\label{sum3}
\Delta X_1^2 \Delta P_1^2\geq \max_{\{|\psi_{n}\rangle\}} \frac{1}{4} \Big(\sum_n \Big|\langle[\bar{X_1},\bar{P_{1 n}^\psi}]\rangle_\psi+\langle\{\bar{X_1},\bar{P_{1 n}^\psi}\}\rangle_\psi\Big|^2\Big),
\end{equation}
where 
\begin{equation}\label{c}
\langle[\bar{X_1},\bar{P_{1 n}^\psi}]\rangle_\psi=\langle\psi|\bar{X_1}|\psi_n\rangle \langle\psi_n|\bar{P_1}|\psi\rangle - \langle\psi|\psi_n\rangle \langle\psi_n|\bar{P_1}\bar{X_1}|\psi\rangle.
\end{equation}
Similarly, the anti-commutation relation follows.

Here, we have considered $|\psi\rangle$ as the state of the system and $\{|\psi_n\rangle\}$ as the basis states of the LHO. We have considered the states of LHO in NC space equivalent to the states of our known commutative space, as the models are developed by coupling the canonical variables of commutating space.

\begin{figure}
  \includegraphics[scale=0.26]{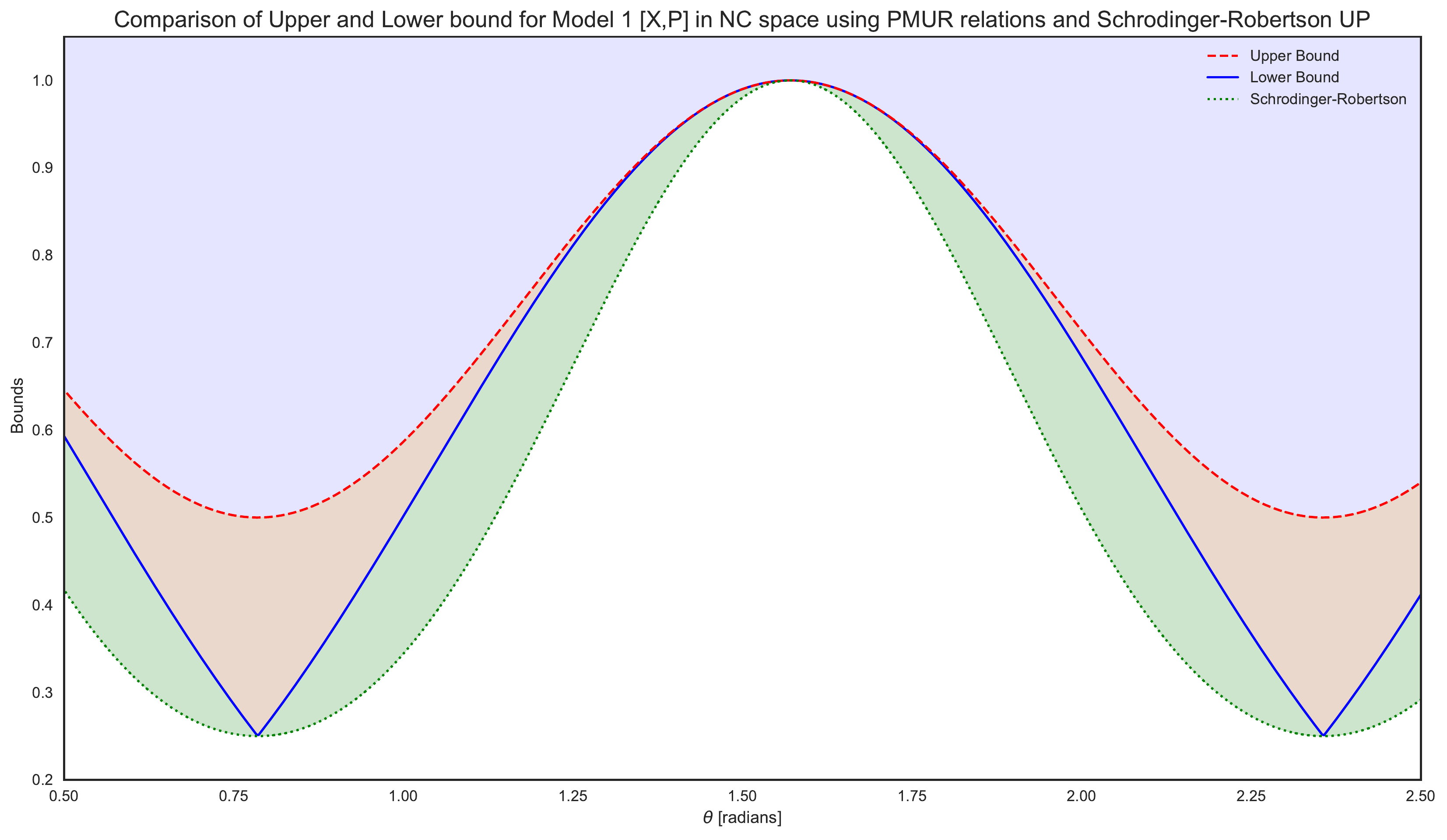}
  \caption{Shown above are the lower (Eq.~\eqref{sum3}) and the upper bound  (Eq.~\eqref{summ2}) for the product of uncertainty of \textit{X} and \textit{P} of \textbf{\textit{Model 1}}, for $|\psi\rangle=cos\theta |\psi_0\rangle-sin\theta |\psi_1\rangle$. Here $\psi_0$ is the ground state and $\psi_1$ is the first excited state of LHO. The green shaded region describes the right side of the SR relation Eq.\eqref{b1}. The doted line is the plot of SR relation. Here it is shown that the lower bound of (Eq.~\eqref{sum3}) is better than Eq.\eqref{b1}. }
  \label{fig1}
\end{figure}

\begin{figure}
  \includegraphics[scale=0.26]{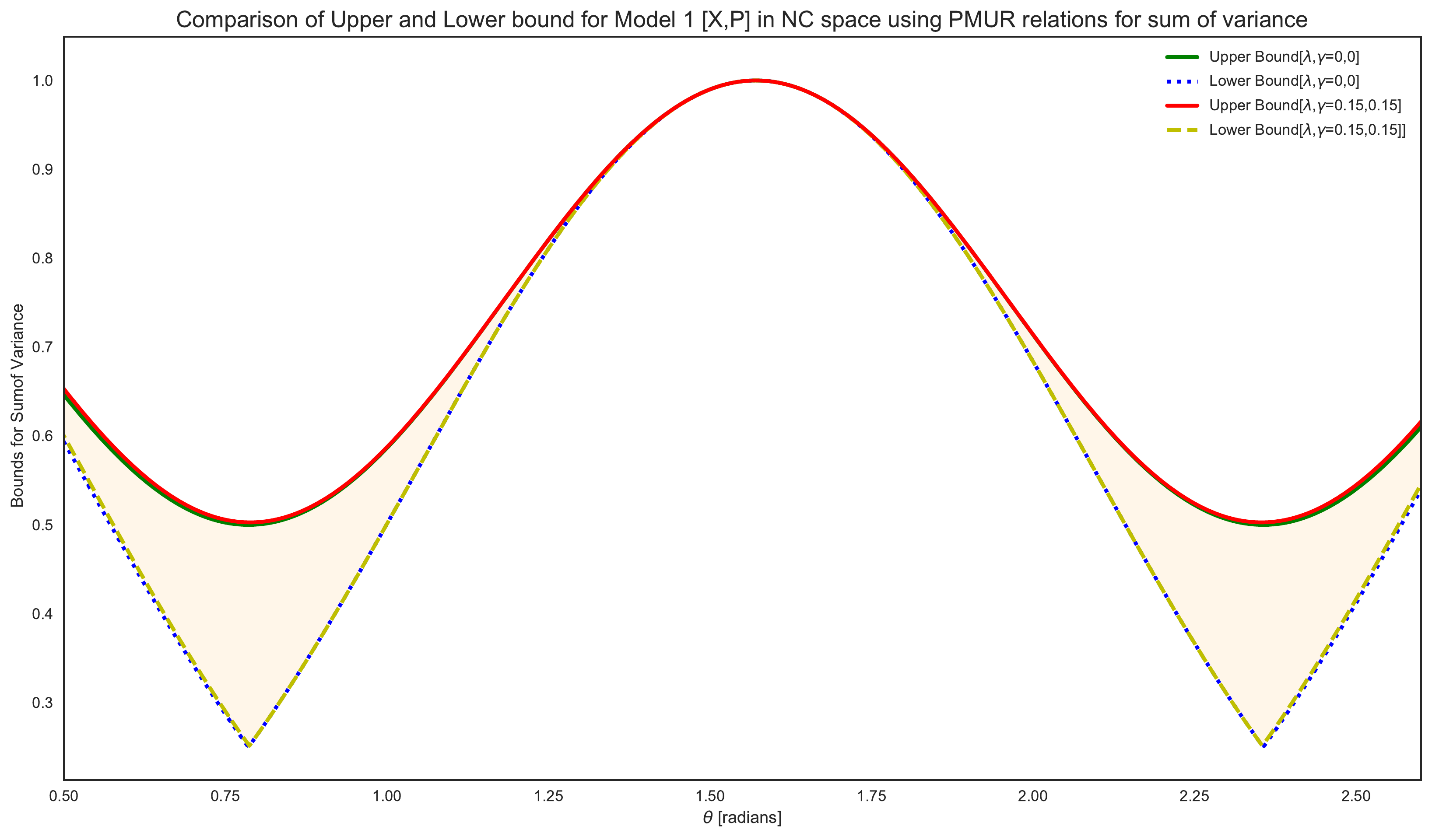}
  \caption{Similar to Fig.~1 we plot for $\lambda=\gamma=0$ and $\lambda=\gamma=0.15$. This plot shows that it is difficult to detect the NC space from the commutating space as the difference is too small.}
  \label{fig2}
  \end{figure}

\textbf{Sum of UR in NC space} is an interesting mathematical object for study. The product of variance can be trivial even for two incompatible observable. This is where  the sum of uncertainty comes into play, where it captures the uncertainty in the observables even when it is non-trivial. Stronger UR  has been put forward before in the work of~\cite{maccone}. Here we are  using the definition proposed by Mondal \textit{et al.} that yields better bounds than the previous ones, without requiring any further optimization.

The Sum of UR for two incompatible observable is 
\begin{equation}\label{o}
\Delta A^2 +\Delta B^2 \geq \frac{1}{2} \sum_n \Big(\Big|\langle\psi_n|\bar{A}|\psi\rangle\Big|+\Big|\langle\psi_n|\bar{B}|\psi\rangle\Big|\Big)^2.
\end{equation}
For the current model,  we replace $A=X_1$ and $B=P_1$, giving
\begin{equation}\label{sum6}
\Delta X_1^2+ \Delta P_1^2 \geq \frac{1}{2} \sum_n \Big(\Big|\langle\psi_n|\bar{X_1}|\psi\rangle\Big|+\Big|\langle\psi_n|\bar{P_1}|\psi\rangle\Big|\Big)^2.
\end{equation}
Similar to Eq.~\eqref{sum6}, we can develop the sum of uncertainty for the position and momentum operators of the NC space equivalently.

\subsection{Model 2: Non-linear model}
In this section, we will present the analysis of a second model, non-linear in nature. One of its application is in string theory~\cite{bala}. The representation of position and momentum operator for this non-linear model are  
\begin{eqnarray}\label{sun22} 
\hat{X_1}=\hat{x_1}, &&  \hat{X_2}=\hat{x_2},\\ \nonumber 
\hat{P_1}=\hat{p_1}(1-\alpha \hat{p_1}+2\alpha^2 \hat{p_1}^2), && \hat{P_2}=\hat{p_2}(1-\alpha \hat{p_2}+2\alpha^2 \hat{p_2}^2),
\end{eqnarray}
where $\alpha=\frac{\alpha_0 l_{pl}}{\hbar}$ and $\alpha_0$ is of order of 1 and
$l_{pl}=10^{-35} m$ (Planck length).

The commutation relation of this non-linear model includes the linear and the quadratic term in $p$, in the Planck regime~\cite{bala}. 
\begin{equation}\label{sun33}
[\hat{X_j},\hat{P_j}]=i\hbar [1-\alpha \hat{p_j}+4\alpha^2 \hat{p_j}^2],	\qquad j=1,2.
\end{equation} 
We are going to develop Eq.~\eqref{sum1} for this current model in a similar fashion.
Evaluating the expectation of the position and the momentum operator, we get
\begin{eqnarray}
\langle\hat{X_1}\rangle\langle\hat{P_1}\rangle &=&\langle\hat{x_1}\rangle\langle\hat{p_1}-\alpha \hat{p_1}^2+2\alpha^2 \hat{p_1}^3\rangle\\ \nonumber
&=&\langle\hat{x_1}\rangle\langle\hat{p_1}\rangle-\alpha \langle\hat{x_1}\rangle\langle\hat{p_1}^2\rangle+2\alpha^2 \langle\hat{x_1}\rangle\langle\hat{p_1}^3\rangle. \nonumber
\end{eqnarray}
The anti commutation relation for this model is given as
\begin{eqnarray}
\{\hat{X_1},\hat{P_1}\}&=&\{\hat{x_1},(\hat{p_1}-\alpha \hat{p_1}^2+2\alpha^2 \hat{p_1}^3)\}\\ \nonumber
&=&\{\hat{x_1},\hat{p_1}\}-\alpha\{\hat{x_1},\hat{p_1}^2\}+2\alpha^2 \{\hat{x_1},\hat{p_1}^3\}. \nonumber
\end{eqnarray}
Plugging in the above relations in Eq.~\eqref{sum1}, we get
\begin{equation}\label{d}
\Delta X_1^2 \Delta P_1^2 \geq \frac{1}{4} \hbar^2 [1-\alpha p+ 4\alpha^2 p^2]^2
 +\frac{1}{4}C^2n[(n-1)^{\frac{1}{2}}+(n+1)^{\frac{1}{2}}]^2,
\end{equation}
where $C=6\alpha^2 \hbar(\frac{\hbar}{2})^{\frac{3}{2}}(m_a\omega)^{\frac{1}{2}}$, $m_a$ is the mass of the particle and $\omega$ is the angular frequency. For Eq.~\eqref{d}, $n$ takes integer values from $[1,\infty]$. The state of the system is having $n$ states, where $n=0$ corresponds to the ground state and $n=1$ correspond to the first excited state and so on.

We will  optimize the UR from the well known form as in Eq.\eqref{sum3}, where one can describe the different components using Eq.~\eqref{c}.

We have encountered an interesting difference while dealing with this linear and non-linear model. For the linear model of NC space and even in the case of commutative space there was no scope of differentiating the Heisenberg and SR relation. But in the case of non-linear model we encountered an extra  scaling factor for SR relation in Eq.~\eqref{d} along with the form that we get while deriving the Heisenberg relation.
 
\begin{figure}
  \includegraphics[scale=0.26]{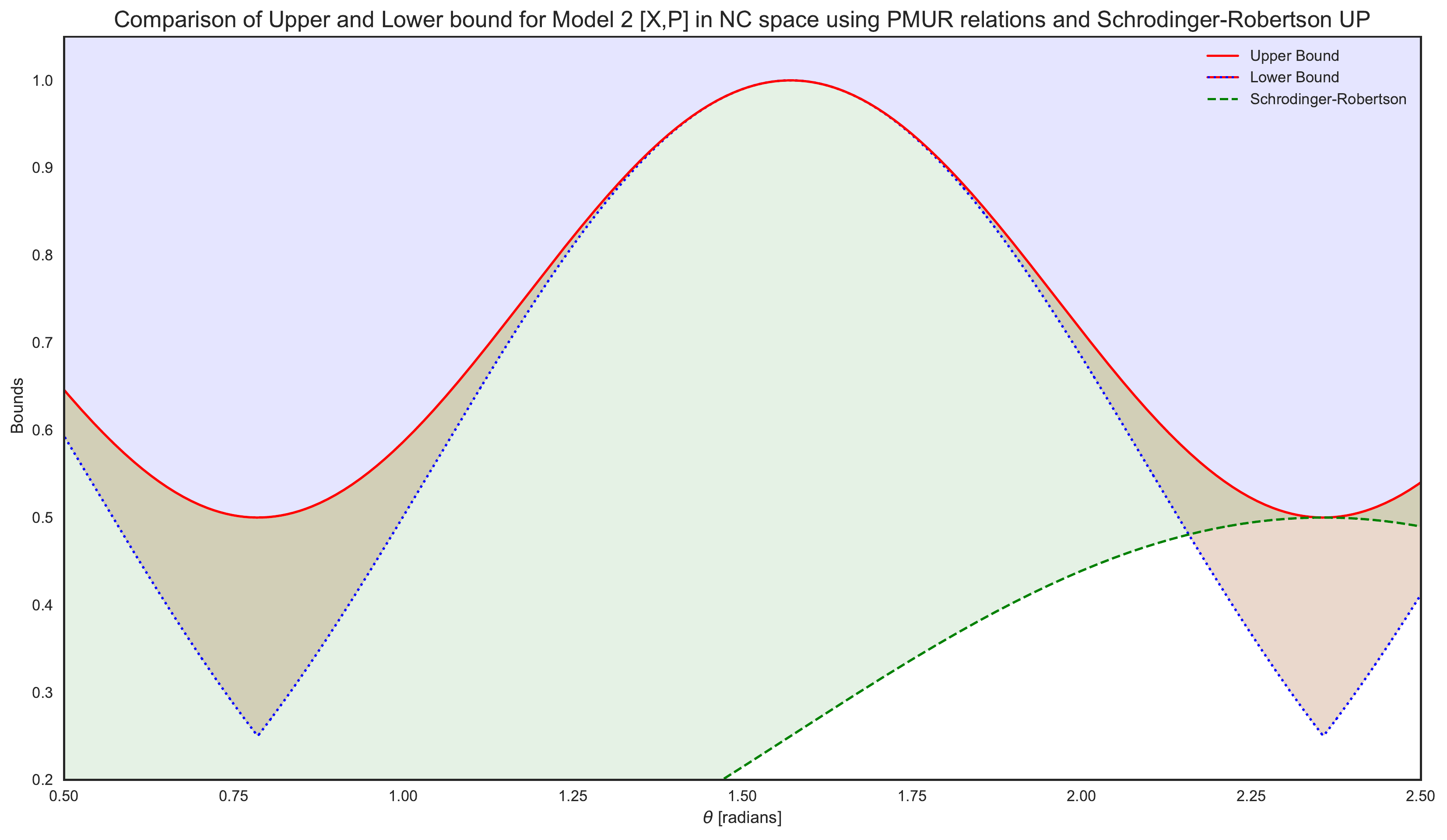}
  \caption{Shown above are the lower (Eq.~\eqref{sum3}) and the upper bound  (Eq.~\eqref{summ2}) for the product of uncertainty of \textit{X} and \textit{P} of \textbf{\textit{Model 2}}, for $|\psi\rangle=cos\theta |\psi_0\rangle-sin\theta |\psi_1\rangle$. Similar to the linear model the lower bound of (Eq.~\eqref{sum3}) is better than the Eq.\eqref{b1} for non-linear model.}
  \label{fig3}
  \end{figure} 

\section{Reverse uncertainty relations for NC space}

 The computation of the reverse UR for  NC space models are presented in this section. It allows to put a constraint on the upper limit of the uncertainty bound. First we are going to develop the upper bound in uncertainty for the linear model followed by the non -linear model. For this we are going to use reverse Cauchy-Schwarz inequality~\cite{rcs,rcs1,rcs2,Dunkl}. It is defined as: 
\begin{equation}
\sum_{i,j} c_i^2 d_j^2 \leq \frac{CD+cd}{4CDcd} \Bigg(\sum_{i,j}c_i d_j\Bigg)^2,
\end{equation}
where $0 < c \leq c_i \leq C < \infty$, $0 < d \leq d_i \leq D <\infty$ for some
constants $c$, $d$, $C$ and $D$ for all $i = 1, ...n$. 
One can obtain the product of variance of two observables using the above inequality as
\begin{equation} \label{summ1}
\Delta A^2 \Delta B^2 \leq \frac{\Lambda_{\alpha\beta}^\psi}{4} \Bigg(\sum_{n}\Big|\langle[~\overline{A},\overline{B}_{n}^{\psi}]\rangle+\langle\{\overline{A},\overline{B}_{n}^{\psi}\}\rangle\Big|\Bigg)^2,
\end{equation}
where $\Lambda_{\alpha\beta}^{\psi\Psi}=\frac{\big(M^{\alpha}_{\psi\Psi}M^{\beta}
_{\psi\Psi}+m^{\alpha}_{\psi\Psi}m^{\beta}_{\psi\Psi}
\big)^2}{4M^{\alpha}_{\psi\Psi}M^{\beta}_{\psi\Psi}m^{\alpha}_{\psi\Psi}m^{\beta}
_{\psi\Psi}}$ with $M^{\alpha}_{\psi\Psi}=\max\{|\alpha_{n}|\}$, $m^{\alpha}_{\psi\Psi}=\min\{|\alpha_{n}|\}$, $M^{\beta}_{\psi\Psi}=\max\{|\beta_{n}|\}$ and $m^{\beta}_{\psi\Psi}=\min\{|\beta_{n}|\}$. Here $\alpha_{n}$, $\beta_{n}$ are the real constants, whose square form represents the probability of finding the particle in that state.
For~\textbf{Model 1}, we have to replace $A=X_1$ and $B=P_1$ in Eq.~\eqref{summ1}, giving  
\begin{equation}\label{summ2}
\Delta X_1^2 \Delta P_1^2 \leq \frac{\Lambda_{\alpha\beta}^\psi}{4} \Bigg(\sum_{n}\Big|\langle[~\overline{X_1},\overline{P_1}_{n}^{\psi}]\rangle+\langle\{\overline{X_1},\overline{P_1}_{n}^{\psi}\}\rangle\Big|\Bigg)^2.
\end{equation}
Unlike the conventional commutative space, one has to develop Eq.~\eqref{summ2} separately for both the position and the momentum operators. This can be generated by substitution of the variables in Eq.~\eqref{summ1} by the position and the momentum operators.
 
 The reverse uncertainty relation for the sum of variance can be developed using the Dunkl-Williams inequality~\cite{Dunkl}. Using this inequality, we get 
 \begin{equation}\label{s}
 \Delta A + \Delta B\leq \frac{\sqrt{2}\Delta (A-B)}{\sqrt{1-\frac{Cov(A,B)}{\Delta A. \Delta B}}}.
 \end{equation}
 Squaring both sides of the Eq.~\eqref{s} we get the upper bound of the sum of variance for two variables as
\begin{equation}\label{sum9}
\Delta A^2 + \Delta B^2 \leq \frac{2 \Delta(A-B)^2}{1-\frac{\textrm{Cov}(A,B)}{\Delta A \Delta B}} - 2 \Delta A \Delta B.
\end{equation}

For our linear model we replace $A=X_1$ and $B=P_1$ of the corresponding Model 1 in Eq.~\eqref{sum9} to get
\begin{equation}\label{b4}
\Delta X_1^2 +\Delta P_1^2 \leq \frac{2 \Delta(X_1-P_1)^2}{1-\frac{Cov(X_1,P_1)}{\Delta X_1 \Delta P_1}}- 2 \Delta X_1 \Delta P_1,
\end{equation} 
where 
\begin{equation}\label{sum10}
Cov(X_1,P_1)=\frac{1}{2}\langle\psi_n|\{X_1,P_1\}|\psi\rangle-\langle\psi_n|X_1|\psi \rangle\langle\psi_n|P_1|\psi\rangle\\
\end{equation}
 and 
\begin{equation}
\Delta (X_1-P_1)^2= \langle \psi_n|(X_1-P_1)^2|\psi\rangle - \langle \psi_n|(X_1-P_1)|\psi\rangle^2.
\end{equation}

 Unlike to the commutative space, to develop the sum of variance for the position operator for our model, we have to consider $A=X_1$ and $B=X_2$ respectively. Similarly, to generate the sum of variance for the momentum operators, we have to substitute $A=P_1$ and $B=P_2$ in Eq.~\eqref{sum9}. The covariance can be calculated in the same fashion as shown in Eq.~\eqref{sum10}.
 \begin{figure}
   \includegraphics[scale=0.26]{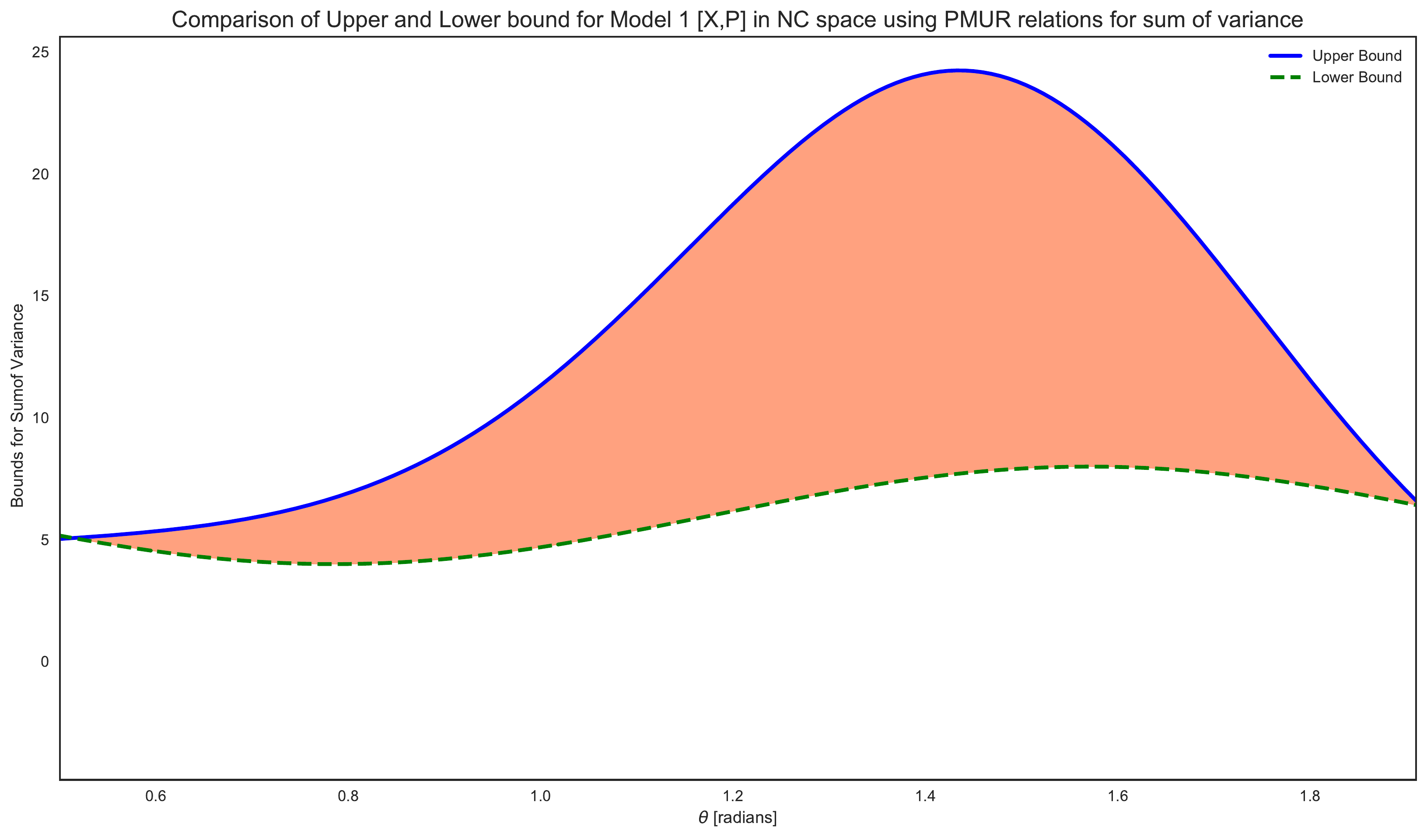}
   \caption{Shown above are the lower (Eq.~\eqref{sum6}) and the upper bound  (Eq.~\eqref{b4}) for the sum of uncertainty of \textit{X} and \textit{P} of \textbf{\textit{Model 1}}, for $|\psi\rangle=cos\theta |\psi_0\rangle-sin\theta |\psi_1\rangle$. This is a general plot with arbitrary (theoretical) values of $\gamma$, $\lambda$.}
   \label{fig4}
   \end{figure}
 
 The treatment for the upper bound for the sum of variances of \textbf{Model 2} is exactly similar to that of the Model 1.  Replacing $A=X_1$ and $B=P_1$ in Eq.~\eqref{sum9} of Model 2, we obtain the  expression of this model in the form, of Eq.~\eqref{b4}.
\begin{figure}
  \includegraphics[scale=0.26]{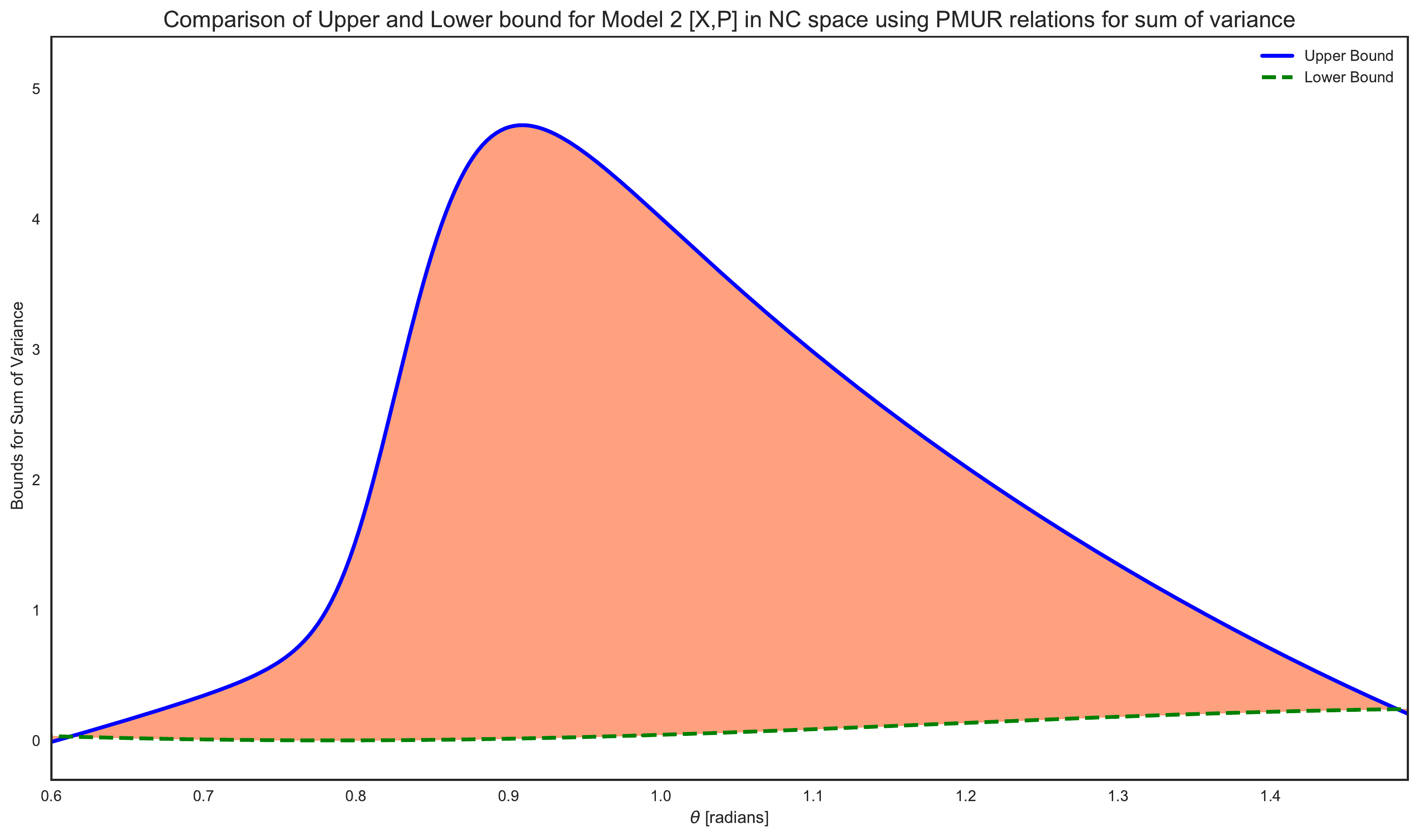}
  \caption{Similar to Fig.~4, but computed with the \textbf{Model 2}. }
  \label{fig5}
     \end{figure}

The tighter bound of the uncertainty relation conveys that for a fixed amount of spread in the measurement outcome of one the observables the amount of spread for the other observable is bounded from both the sides. Experimental realization to probe deformations of the canonical commutator~\cite{pik} and non-
commutative theories~\cite{sdey} using quantum optics have been explored. So, the bound in the uncertainty measure in non-commutative space can be experimentally verified by extending the approach followed in the work~\cite{qmref}.
     
 The bound in the uncertainty will be an important factor in the quantum metrology in non-commutative space structure. It will pose an upper bound in the error of the measurement and quantum evolution.

\section{Discussion and Conclusion}
 To summarize our results, we see from the plots, that our computation of the LHO using NC models are consistent yet not similar to the results obtained from the traditional commutative space models. From Fig.~1, we can infer that the lower bound(blue curve) is better than the SR relation(green dotted curve). In Fig.~2,  we have re-plotted Fig.~1 (excluding the SR relation), but for two different set of parameter values ($\gamma$, $\lambda$, [$0,0.15$]). Where the first case $\gamma,\lambda=0$, is a special case where the model reduces to the standard commutative space case. Here we can  see that the difference between the bounds of the commutative space($\gamma,\lambda=0$) and the non-commutative space($\gamma,\lambda=0.15$) is very small. We speculate that, this  is why differentiating between the commutative and NC space is practically challenging. Fig.~3 is generated with the non-linear model. This plot further verifies the consistency of the PMUR relations in the NC space. However owing to the non-linear nature of this model the curve from the SR relation appears to have deviation from the nature exhibited in the Fig.~1 and~2. We can also see that in a small part the SR relation exceeds the lower bound provided by the PMUR in this model (shaded drab region), thus making it open for speculation on the  tightness of the bound over the SR lower bound in such non-linear NC models.  In the Fig.~4 and~5 we have presented the upper and the lower bound from the sum of variances using the PMUR relations. The plot gives us an allowed region for the range of uncertainty(for PMUR using sum of variances only) to be valid. One can see that the linear model (Fig.~4) gives a less stricter/more wider allowed range(shaded orange region) for the PMUR relation in comparison to the Model 2, which is non-linear. The reason for such difference between the models is open for further speculations. Future experimental verifications of this results could help the community better with the understanding of the nature of working of these two models. In addition from Eq.~\eqref{d}, we have shown to the best of our knowledge for the first time that there is difference in bounds between the SR relation and the Heisenberg relation for a non-linear model operating in NC space. The URs have been the corner stone of quantum theory. Even after nine decades of  evolution of the URs, it is still open for further analysis and speculations.

To summarize, we have established the tighter URs for a linear and a non-linear model for two incompatible variables in NC space. We have also established the upper bound of the UR for the sum and product of our models. Together this URs and the bounds can play an important role in quantum cryptography and quantum metrology. For example, depending on the error-margin of the underlying space, one can select an error-correcting code to design error-free algorithm and protocols in that space. This can potentially lead to optimization of error analysis in quantum domain. 

The impact of the bounds of the uncertainty relation in quantum thermodynamics is also worth studying. Even one can apply UR to find bounds on thermodynamic variables.

The physical implication and the affect of the upper bound of the uncertainty relation on quantum field theory can be explored. The models we have studied are non-canonical in form. The results are similar to the canonical structure which is quite surprising and needs further study.

\section{Acknowledgments}
We would like to thank Prof. Subir Ghosh of Physics and Applied Mathematics Unit, Indian Statistical Institute, Kolkata, for his valuable inputs and suggestions.



\end{document}